\documentclass[useams,usenatbib]{mn2e}
\usepackage[dvips]{graphicx}
\usepackage{times}
\title[Hot Cores : Probes of High-Redshift Galaxies?]{Hot Cores : Probes of High-Redshift Galaxies?}
\author[Lintott et al.]{C.J.Lintott$\thanks{Email : cjl@star.ucl.ac.uk}$, S.Viti, D.A.Williams, J.M.C.Rawlings, I.Ferreras \\
Department of Physics and Astronomy, University College London, Gower Street, London, WC1E 6BT, UK}
\begin{document}
\date{March 2005}

\pagerange{\pageref{firstpage}--\pageref{lastpage}} \pubyear{2005}

\maketitle

\label{firstpage}

\begin{abstract}

The very high rates of second generation star formation detected and
inferred in high redshift objects should be accompanied by intense
millimetre-wave emission from hot core molecules. We calculate the
molecular abundances likely to arise in hot cores associated with 
massive star formation at high redshift, using several different models of
metallicity in the early Universe. If the number of hot cores exceeds 
that in the Milky Way Galaxy by a factor of at least one thousand, then a 
wide range of molecules in high redshift hot cores should have detectable
emission. It should be possible to distinguish between different models 
for the production of metals and hence hot core molecules should be useful probes of 
star formation at high redshift.

\end{abstract}


\section{Introduction}

In recent years there has been much debate - both observational and theoretical - about star formation in the early Universe. For example, strong evidence seems to point to a greatly enhanced star formation rate earlier in the evolution of the Universe (\citet{Calzetti}, but see also \citet{Bunker}). Individual objects with greatly enhanced star formation rates have also been detected; a good example is the quasar at a redshift of 6.4 studied by \citet{Bertoldi}. They detect CO and, via the far-infrared emission, they derive a star formation rate of $3000 \mathrm{M_\odot yr^{-1}}$.

Simulations such as those carried out by \citet{Abel} and \citet{Bromm} (see also \citet {BrommLarson} for a review of recent work in this field) have investigated the nature of the first stars, which are expected to have formed at a redshift in excess of 10. Such work indicates that these first stars were large, with masses in excess of 100$\mathrm{M_\odot}$. These massive stars must have a profound influence on the environment around them; they may be the cause of the reionization of the Universe \citep{BarkanaLoeb}, and the metals produced by them may seed the interstellar medium for later star formation. Hence second generation stars should form after the first from gas already enriched with metals.

If such large rates of star formation are present in the early universe, they should be observable. Molecular signatures are distinct and powerful indicators of star-forming activity in our own galaxy; in this paper we seek to investigate their visibility and utility in systems at high redshift.

Specifically, we consider the possibility of observing one particular stage of this star-formation; the 'hot core' phase. This is a short lived episode in the evolution of a high-mass protostar (although recent work \citep{Bottinelli} has shown that low-mass stars too pass through a similar phase). During the initial collapse of the forming star, many molecular species which are present in the gas phase 'freeze out' onto the surface of dust grains. This process has a profound influence on the chemical evolution of the core: Firstly, potential reactants and coolants are removed from the gas phase, and secondly the grain surfaces themselves provide a site for enhanced further chemical evolution, producing more complex species than would otherwise be possible. Once the newly-formed star has begun to radiate, heating the surrounding gas, the ice sublimates and the newly formed (generally more complex) molecules are returned to the gas phase. These newly released molecules radiate strongly in the warm gas. This is the hot core phase; it is this period in the protostar's evolution with which we are concerned in this paper.

Despite much interest in star formation at high redshift, there has been no detailed study of the possible contribution of chemical models of the type used in studies of nearby star forming regions to observations. The question addressed here is: Will the molecular emission associated with 'hot cores' be detectable in distant (high-redshift) systems?

\section{Methodology}
\subsection{Initial Abundances}

In adapting extant chemical models to reflect the situation at high redshift, the selection of initial abundances for atomic species is crucial. In seeking to study the formation of a second generation of stars, we use models predicting the yield from the massive first generation. These yields will critically depend on the mass of the progenitor; for example, it has been argued that zero-metallicity stars with a mass in the range 40-140 $\mathrm{M_\odot}$ collapse directly to a black hole with little mixing with the surrounding medium \citep{Heger03}.

For this initial study we use three different models of the yields from a zero-metallicity star. 
\citet{CL} (hereafter CL02) have performed calculations using the FRANEC code \citep{FRANEC} to predict the results from a set of zero metallicity stars with initial masses between 15 and 80 $\mathrm{M}_\odot$. They perform a detailed comparison between their models and the 'average' observation obtained by Norris et al. (2001) from five low-metallicity stars. We use the best fit model for a progenitor with a mass of 80 $\mathrm{M}_\odot$ as the starting point for our calculations. However, it is obvious that the yield from such an event would be diluted by mixing with the pristine surrounding gas, and so we adopted specific values of the carbon abundance as a fraction, $f$, of local abundance \citep{Sofia} and then matched the abundance ratio X/C (for any atomic species X) to the model in CL02.

We have also considered the models of \citet{UN} (hereafter UN02), who use a separate set of codes to model progenitors of pair instability supernov\ae with masses from 150 to 270 $\mathrm{M}_\odot$.  In a supernova of this type, the creation and annihilation of positron-electron pairs causes the core to be so unstable that gravitational collapse is halted. They are therefore particularly interesting in this context as they may produce no dense remnant, scattering all of their material into the interstellar medium.

Finally, we use abundances based on the models of \citet{HW} (hereafter HW02) for a star with a 80$\mathrm{M_\odot}$ helium core, corresponding to a total mass of 200$\mathrm{M_\odot}$. The initial abundances relative to carbon used for each of the three models are shown in table \ref{table:snabund}. Three different models were included in order to (a) investigate the sensitivity of our results to the choice of enrichment model and (b) investigate the possibility that future observations could distinguish between different models.

It should be noted that the nitrogen abundances constitute the largest differences between the models. The synthesis of
primary nitrogen is not considered in detail by the current
supernova models. For instance, HW02 only evolve helium cores, thereby neglecting the (significant)
production of nitrogen during the earlier stellar phases.
Furthermore, the nitrogen yields are highly sensitive
to the poorly constrained mechanisms of rotation and
convective overshoot. Nitrogen is mostly a secondary element
as illustrated by the high abundance ratio found in the Sun.

\begin{table}
\begin{tabular}{|c|c|c|c|c|}
\hline
Species&CL02&UN02&HW02&Solar\\
\hline
C&1.0000&1.0000&1.0000&1.0000\\
O&3.212&8.3887&10.833&2.719\\
N&4.240(-7)&0.0023&1.1155(-5)&0.293\\
S&0.05375&0.2874&1.0895&0.172\\
Mg&0.1293&0.4373&0.8504&0.232\\
\hline
\end{tabular}
\caption{Abundances relative to carbon, obtained from the supernova models of \citet{CL}, \citet{UN} \& \citet{HW}. CL02 assume a progenitor of 80 $\mathrm{M}_\odot$, whereas UN02 model a larger star with a mass of 150 $\mathrm{M}_\odot$. HW02 use a model which produces a total yield of 80 $\mathrm{M_\odot}$. These abundances provide the initial conditions for the results presented in section \ref{sec:results}. The largest difference is the near-absence of nitrogen in the CL02 and HW02 models. Solar abundances are from Grevesse \& Sauval (1998).}\label{table:snabund}
\end{table}

\subsection{Chemical and Dynamical Model}

We have used a chemical model, similar to that described in Viti \& Williams (1999), to describe the chemical evolution of regions of high mass star formation. The model is essentially a two stage calculation: the first stage begins with a diffuse ($\sim$ 300 cm$^{-3}$) medium in purely atomic form (apart from a fraction of hydrogen incorporated into $\mathrm{H_2}$) and then undergoes a collapse under free-fall. The dynamical and chemical equations are solved simultaneously during this phase. We use the free-fall collapse law described in Rawlings et al. (1992) in which the change in central gas number density, $n$, as collapse proceeds from initial density $n_0$ to final density $n_f$ is given by 

\begin{equation}
\frac{\mathrm{d}n}{\mathrm{d}t}  = \left(\frac{n^4}{n_0}\right)^{\frac{1}{3}}\left( 24\pi Gm_{H} n_{0} \left[ \left(\frac{n}{n_0}\right)^{\frac{1}{3}}-1\right]\right)^{\frac{1}{2}}
\end{equation}

It can be seen from this equation that we do not attempt to include the internal structure of the core in our calculations. During this initial collapse phase the temperature is 10K. Once a critical density (chosen to represent observed densities of hot cores) is reached the collapse stops, the temperature increases to 300K and the system is allowed to continue evolving chemically, although the dynamical evolution has stopped. For massive stars, the increase in temperature due to the ignition of the protostar occurs extremely rapidly, justifying our choice of an instantaneous increase in temperature. 

The chemical network we use is taken from the UMIST rate 99 database \footnote{www.rate99.co.uk} \citep{UMIST}, and we follow the chemical evolution of 168 species involved in 1857 gas-phase and grain reactions. During the first phase, all species except hydrogen and helium freeze onto the grains and, where possible, hydrogenate. At the end of the first stage, approximately 99 percent of these species has frozen onto the grains \citep{Caselli}.

\subsection{Dust Properties \& Extinction}

We have considered the effects of the differences in the metallicity of infalling material between low and high redshift star formation, and have also attempted to allow for the low abundance of dust available in the early Universe. The formation and evolution of dust in the early Universe is extremely poorly understood. We do not know the chemical nature of dust, nor the dust grain size distribution and therefore we do not know the appropriate interstellar extinction curve to use. It should be clear from the description of the hot core phase of protostellar evolution earlier that these issues may prove to be crucial. However, in the absence of data which might allow us to make more informed decisions we have simply assumed that the dust mass scales with metallicity and that the size distribution generates an interstellar extinction curve similar to the galactic results. 

Another important factor could be the evolution of the cosmic microwave background (CMB). The evolution of the temperature of the blackbody spectrum is given by $T=T_0\left(1+z\right)$ and the well-known Planck formula then gives the intensity for a particular wavelength at a particular time. 

The important wavelengths - those which would participate in the dissociation of molecules - are in the ultraviolet. Even at a redshift of 10, the order of magnitude increase in CMB temperature is accompanied by an insignificant increase in intensity at these wavelengths. In addition, it should be noted that at z=6 the temperature of the CMB would be about 20K, and this may begin to affect the freeze-out of species necessary for the formation of hot-core chemistries we have discussed here. However, as long as the cores being considered remain optically thick then the effect should be minimal.

In local hot cores, the extinction is usually estimated to be between a few hundred and a few thousand visual magnitudes (see, for example, van der Tak et al. 2000). The extinction of the core will obviously be reduced at higher redshifts due to the lack of the dusty component. If the accompanying drop in visual extinction is severe enough then, as the cloud becomes optically thin, photo-dissociation will destroy the molecules associated with a hot core chemistry. In the case already considered of dust at one thousandth of its density in the Milky Way Galaxy, increasing the diameter of the core by a factor of 10 (to 0.3pc) produces an extinction of 10 to 15 magnitudes, which is equivalent to sufficient optical depth. We have carried out our calculations for cores of this size.

The survival of the molecules is, nonetheless, uncertain. However, the detection of CO at a redshift of 6.42 by Walter et al.(2003) provides evidence that molecules can survive in these early epochs. It should also be noted that the change in the spectrum of the CMB will alter the excitation of molecular species (Redman et al, 2004).

\section{Results}\label{sec:results}
\subsection{Single Hot Core}

The results of the initial runs, expressed as column densities, are shown in tables \ref{table:cl} and \ref{table:un}. There are differences between the models, regardless of the overall metallicity adopted. In particular, the UN02 150$\mathrm{M}_\odot$ models yield much higher predicted column densities for sulfur-bearing species such as CS, $\mathrm{H_2CS}$ and $\mathrm{SO_2}$.

The effects of the changes to size and hence $A_{\mathrm{v}}$ discussed above are clearly shown in table \ref{table:cl}. Although the general trend is for the column density to be reduced, the effect is small and for some species the reduction in $A_{\mathrm{v}}$ when coupled with the increase in size is an increase in column density.

Chieffi \& Limongi have recently published new calculations for the yields of smaller stars with zero or low metallicity \citep{CL04}. We find that runs which use the yields from a zero metallicity star of mass 30$\mathrm{M_\odot}$ produce column densities approximately an order of magnitude larger than those presented in table \ref{table:cl}. A further enhancement is found if the yields from a star with metallicity $10^{-4}Z_{\odot}$ are used as initial conditions. \footnote{Those interested can obtain tables in the same format as table 2 for these runs at http://www.star.ucl.ac.uk/~cjl/CL04tables.html}

\begin{table}
\begin{tabular}{|l|c|c|c|c|}
\hline
Species&SN,1/100&SN,1/500&SN,1/1000&*\\
\hline
CO&6.5(15)&1.8(15)&1.1(15)&1.1(15)\\
$\mathrm{H_2CO}$&2.9(14)&9.3(13)&4.2(13)&4.4(13)\\
$\mathrm{H_2S}$&1.5(9)&2.3(9)&3.9(9)&4.1(9)\\
CS&1.2(15)&5.7(14)&3.1(14)&3.1(14)\\
$\mathrm{H_2CS}$&6.9(14)&1.1(14)&3.4(13)&3.7(13)\\
SO&7.9(14)&7.8(12)&2.7(12)&2.8(12)\\
$\mathrm{SO_2}$&7.1(14)&2.1(12)&3.7(11)&4.3(11)\\
$\mathrm{CH_3OH}$&1.1(12)&3.1(10)&9.6(9)&9.4(9)\\
$\mathrm{HC_3N}$&2.5(12)&3.6(11)&1.1(11)&1.1(11)\\
$\mathrm{CH_3CN}$&3.1(9)&1.9(9)&1.2(9)&1.2(9)\\
HCN&7.6(10)&1.1(11)&3.9(11)&2.7(11)\\
HNC&2.9(10)&3.7(10)&1.3(11)&8.7(10)\\
HCO+&5.1(9)&5.6(9)&1.7(10)&8.9(9)\\
\hline
\end{tabular}
\caption{Calculated column densities (in $\mathrm{cm^{-2}}$) for a variety of species; first three columns have a carbon abundance set to 1/100, 1/500 and 1/1000 of solar, and the other species set so that their ratio with carbon matches that predicted by a model for a zero metallicity star which yields 80$\mathrm{M_\odot}$ . (This model is from CL02). The results in the fourth column show the effect of using a model with a carbon abundance of 1/1000 solar but with no change either to the diameter or to the extinction compared to standard models appropriate to the Milky Way. We define 1.2(15) = 1.2 x $10^{15}$}\label{table:cl}
\end{table}

\begin{table}
\begin{tabular}{|l|c|c|c|}
\hline
Species&SN,1/100&SN,1/500&SN,1/1000\\
\hline
CO&6.0(15)&3.5(15)&2.2(15)\\
$\mathrm{H_2CO}$&1.0(14)&3.2(13)&6.2(12)\\     
$\mathrm{H_2S}$&9.6(9)&5.4(9)&7.6(9)\\ 
CS&2.7(15)&1.9(15)&1.1(15)\\
$\mathrm{H_2CS}$&1.8(15)&5.4(14)&1.3(14)\\
SO&1.1(15)&4.0(12)&2.7(12)\\
$\mathrm{SO_2}$&1.1(15)&6.7(11)&2.1(11)\\
$\mathrm{CH_3OH}$&1.3(12)&9.7(10)&2.4(10)\\
$\mathrm{HC_3N}$&4.4(12)&3.6(12)&2.5(12)\\
$\mathrm{CH_3CN}$&4.7(9)&4.3(10)&7.8(10)\\
HCN&1.4(12)&2.8(10)&2.5(12)\\
HNC&3.1(9)&8.1(9)&7.4(9)\\
HCO+&1.7(9)&3.9(9)&5.4(9)\\
\hline
\end{tabular}
\caption{As table \ref{table:cl}, except for yields from a 150 solar mass supernova as calculated by UN02}\label{table:un}
\end{table}

\begin{table}
\begin{tabular}{|l|c|c|c|}
\hline
Species&SN,1/100&SN,1/500&SN,1/1000\\
CO&1.4(16)&3.3(15)&1.8(15)\\
$\mathrm{H_2CO}$&3.1(12)&5.4(11)&6.1(10)\\     
$\mathrm{H_2S}$&7.6(11)&2.5(11)&1.6(11)\\ 
CS&3.3(15)&3.2(15)&2.1(15)\\
$\mathrm{H_2CS}$&4.1(15)&1.3(15)&3.1(14)\\
SO&1.4(16)&7.8(14)&2.2(14)\\
$\mathrm{SO_2}$&9.3(15)&1.3(14)&1.9(13)\\
$\mathrm{CH_3OH}$&4.7(12)&9.2(10)&2.4(10)\\
$\mathrm{HC_3N}$&5.0(7)&3.2(8)&2.6(9)\\
$\mathrm{CH_3CN}$&1.9(6)&2.9(7)&7.8(7)\\
$\mathrm{HCN}$&1.4(10)&2.8(10)&2.5(10)\\
$\mathrm{HNC}$&3.1(9)&8.1(9)&7.4(9)\\
$\mathrm{HCO+}$&1.65(9)&3.9(9)&5.4(9)\\
\hline
\end{tabular}
\caption{As table \ref{table:cl}, except for yields from models by HW02.}\label{table:hw}
\end{table}

\subsection{Detectability of Multiple Hot Cores}

Will these molecular abundances generate detectable signals from high redshift galaxies? In order to answer this question it is necessary to estimate the number of hot cores in a system which might be forming a second generation of stars for us to observe. 

Lumsden et al. (2002) detect 3000 massive young stellar objects in the plane of our galaxy. Making a reasonable adjustment for the incompleteness of their survey we obtain a conservative estimate of $\sim 10^4$ hot cores in our own Milky Way, which has a star formation rate of $\sim 1 \mathrm{M}_\odot\mathrm{yr}^{-1}$. \citet{Bertoldi} estimate from far-infrared luminosity that QSOs of the type considered here may have a star formation rate as high as $\sim 3000 \mathrm{M}_\odot \mathrm{yr}^{-1}$. If the number of hot cores is considered to be proportional to the overall star formation rate, this might suggest that there could be $\sim 10^7$ to $\sim 10^8$ hot cores in such a system.

Is it realistic to expect this many hot cores at high redshifts? \citet{Carilli} report a detection of CO in a QSO at a redshift of z=4.69 using the Very Large Array. We calculate (see Appendix A), considering beam dilution effects, that the signal strength from $1.4\mathrm{x}10^7$ hot cores at a redshift of z=4.69 would produce a signal of the same strength as the Carilli et al. CO result. This assumes that the observations were of transitions in molecules which have a column density at least equal to that of CO in the Carilli et al. source. It should be noted that this analysis makes no identification of the CO detected by Carilli et al. as CO present in hot cores; we merely use the extant observations as a guide to what is detectable.

Alternatively, \citet{Bertoldi} report detection of CO from a QSO at  redshift of z=6.42. They do not identify extended emission at a resolution of 1.5 arcseconds. Repeating our calculation at this redshift, assuming a source size of 1.5 arcseconds (the worst-case scenario) we find that one requires $2.5\mathrm{x}10^8$ hot cores to provide the same signal level. Assuming a source size of 0.3 arcsec, the size of the source considered by Carilli et al., we find $1.01\mathrm{x}10^7$ hot cores to be sufficient.

When this many hot cores are present in an unresolved high redshift source, many molecular species reach significant column densities. Once again, we note that there are differences between the three models, regardless of the overall metallicity adopted. In particular, the UN02 150$\mathrm{M}_\odot$ models have much higher predicted column densities for sulfur-bearing species such as CS, $\mathrm{H_2CS}$ and $\mathrm{SO_2}$. Other species such as $\mathrm{CH_3CN}$ and $\mathrm{CH_3OH}$ are more sensitive to the overall metallicity than the selection of model.

\section{Conclusions}

We have demonstrated that enhanced levels of star formation expected in systems incorporating QSOs at high redshifts provide an opportunity to investigate the nature of the first generation of stars. Systems containing $\sim 10^7$ hot cores should produce a signal at the same level as existing detections of molecular CO in these systems. (Note that we do not claim that existing detections of molecular CO are signatures of hot cores.). Furthermore, our detailed chemical modelling of such systems suggests that a variety of molecular \textbf{species, which} have lines accessible by facilities such as the Very Large Array, have sufficient column density to be detectable. An example is SO, for which molecule the J=5-4 transition is detectable by the VLA for redshifts in the range $4.50>\mathrm{z}>3.40$ and the J=8-7 transition for the range $7.66>\mathrm{z}>5.93$.

We also show, by considering the change in the model predictions produced by changing the initial abundances, that this type of analysis and observation could distinguish between competing models for the first generation of stars. The yields from the supernov\ae  which must end the lives of these stars form the input to our model of the formation of the next generation. As an example,we show that we can distinguish between the model by CL02 and that for UN02 by, for example, comparing the behaviour of SO with that of CO in each model.

The importance of this work is not limited to searches specifically aimed at the formation of the second generation of stars. Indeed, this should be regarded as a 'worst-case' scenario for the potential of searches for high redshift molecules. Later star formation would only further enhance the proportion of metals available for molecule formation, increasing the chance of success. The runs using yields from \citet{CL04} suggest that our prediction of significant column densities holds even if the first generation of stars are not as massive as Abel et al. predict. Nor should we be limited in choice of targets; Ly-break galaxies may prove worth investigation, but more detailed models of chemical enrichment (similar to that in Matteucci and Calura, 2005) are needed to make specific predictions. 

In conclusion, we predict that observational searches for the multiple molecular species associated with the hot core stage of high-mass star formation at high-redshift would be likely to succeed, given the extant prediction of very high star formation rates in objects in this redshift range. We also demonstrate through this preliminary work that the use of sophisticated chemical models could be used in association with observations of this nature to provide constraints on models for the first stars. In the foreseeable future, facilities such as ALMA and the SKA will provide a huge increase in our capacity to carry out this kind of analysis, but our work shows that important results could be obtained with existing facilities.

\section*{Acknowledgements}
C.J.L. is supported by a PPARC studentship. S.V. acknowledges individual
financial support from a PPARC Advanced Fellowship. D.A.W. thanks the
Leverhulme Trust for the Leverhulme Emeritus Award. The authors would like to thank the referee, Prof. Neal Evans, for his helpful contribution.

\section*{Appendix A : Estimating the Number of Hot Cores Required}

First, we must calculate the filling fraction of a single hot core; assuming the hot core has a size of 0.03pc. At z=4.69, the angular diameter distance is 1.3361Gpc, and so this corresponds to an angular diameter of

$\theta_{hc}=2.25 \mathrm{x} 10^{-6}$ arcsec

and therefore a filling fraction

$f=\frac{\theta_{hc}^2}{\theta_{hc}^2 + \theta_{Beam}^2}=5.96 \mathrm{x} 10^{-15}$

$\theta_{Beam}$ is the beam size, which for the VLA in the Q band (40-50 GHz) is 1'. 

Adjusting for the temperature of the hot core, we multiply by a factor of T=300 K.

This is then set to be equal to the strength of the CO detected by Carilli et al. Their source had an apparant radius of 0.3 arcsec. Hence, the number of hot cores, n, required for a detection at the CO strength is :

$nTf=f_{CO}=\frac{0.3^2}{0.3^2+60^2}$

For this source $n=1.40 \mathrm{x} 10^7$

Similar arguments can be used at higher redshifts. Walter et al. (2003) present observations of a quasar at a redshift of 6.42, and do not resolve the CO emission at a scale of 1.5 arcsec; hence using the Carilli et al. size of 0.3 arcsec gives 

$n=1.01 \mathrm{x} 10^7$

while using 1.5 arcsec gives

$n=2.5 \mathrm{x} 10^8$

\label{lastpage}

\end{document}